\documentclass[aps,prd,twocolumn,superscriptaddress,nofootinbib,preprintnumbers]{revtex4}

\usepackage[pdftex,colorlinks]{hyperref}
\usepackage[dvipsnames]{xcolor}
\definecolor{phthaloblue}{rgb}{0.0, 0.06, 0.54}

\hypersetup{
    colorlinks=true,
    linkcolor=MidnightBlue,
    citecolor=MidnightBlue,
    filecolor=MidnightBlue,
    urlcolor=MidnightBlue,
    }
\usepackage[pdftex]{graphicx}
\usepackage{bm}
\usepackage{cancel}
\usepackage{here}
\usepackage{comment,braket}
\usepackage{epsf}
\usepackage{amsmath}
\usepackage{graphics}
\usepackage{amsfonts}
\usepackage{amssymb}
\usepackage{latexsym}
\usepackage{natbib}
\usepackage[normalem]{ulem}
\input{colordvi.tex}
\usepackage{feynmp}
\usepackage[utf8]{inputenc}
\DeclareUnicodeCharacter{2212}{-}
\usepackage{orcidlink}
\usepackage{cleveref}
\crefname{section}{Sec.}{Sec.}
\crefformat{section}{Sec.~#2#1#3}
\crefmultiformat{section}{Sec.~#2#1#3}{-#2#1#3}{, #2#1#3}{-#2#1#3}
\crefname{equation}{Eq.}{Eqs.}
\crefrangelabelformat{equation}{(#3#1#4--#5#2#6)}

\preprint{\tt FERMILAB-PUB-26-0209-T}

\begin{document}

\title{Hiding in the Shadow of the Upsilon: \\
Ditaus from a Light Pseudoscalar}

\author{Matthew R.~Buckley}
\affiliation{NHETC, Department of Physics and Astronomy, Rutgers, Piscataway, NJ 08854, USA}

\author{David Shih}
\affiliation{NHETC, Department of Physics and Astronomy, Rutgers, Piscataway, NJ 08854, USA}

\author{Isaac R.~Wang}
\affiliation{Theory Division, Fermi National Accelerator Laboratory, Batavia, IL 60510, USA}
\begin{abstract}
The CMS collaboration has reported a measurement of $\Upsilon$ decays to ditaus using $61.9~{\rm fb}^{-1}$ of scouting data. If interpreted as the decay of $\Upsilon(1S,2S,3S)$, the measured ditau rate is more than ten times that seen in the dimuon final states at the $\sim 3 \sigma$ level, and is likewise inconsistent with the branching ratios measured at $B$-factories. If confirmed with more data and at higher significance, such a violation of lepton flavor universality would necessitate new physics. In this Letter, we present a simple model with a light pseudoscalar coincidentally near the $\Upsilon(1S)$ mass, which mixes with two Higgs doublets in the alignment limit. Such a particle naturally decays primarily to taus and evades all existing experimental constraints, while implying a number of predictions that can be tested in the near future.
\end{abstract}

\maketitle

{\bf Introduction.}
While the Large Hadron Collider (LHC) program continues to push the energy frontier, data scouting techniques have enabled lower trigger thresholds, opening a new window on low invariant mass processes in the presence of large backgrounds.
Recently, with $61.9~{\rm fb}^{-1}$ of $\sqrt{s} = 13.6$~TeV Run 3 scouting data taken during 2022 and 2023, the CMS collaboration has announced the first-ever measurement of the $\Upsilon$ decay to ditau final states at a hadron collider \cite{CMS:2026mwx}.  This tour de force analysis combined novel scouting triggers with innovative machine learning techniques in order to access the difficult phase space at low invariant mass and tau $p_T$.
The success of these techniques demonstrates the potential for new physics searches in the third generation that can be expected in the full set of available scouting data. 

Interestingly, in the  61.9~fb$^{-1}$ dataset the number of ditau events near the $\Upsilon$ mass is found to be unexpectedly large when compared to the recently measured dimuon channel \cite{CMS:2026ccg}. 
Due to the presence of neutrinos in the tau decays, the full invariant mass of the resonance responsible for the large ditau rate cannot be completely reconstructed, and the individual $\Upsilon(1S)$, $\Upsilon(2S)$, and $\Upsilon(3S)$ states cannot be resolved. If the measured ditaus are interpreted as the combination of all three $\Upsilon\to \tau^+\tau^-$ channels, the production cross section times branching ratio is found to be $3.5\pm 0.7 ({\rm stat.}) \pm 0.7 ({\rm sys.})$~nb \cite{CMS:2026mwx}, consistent across multiple $\tau$ decay channels (3-prong+1-prong and 1-prong+strips). This is a factor of 17 times larger than the $\sim 0.2$~nb expectation for these three states given the dimuon measurement in the same fiducial region of $p_T(\Upsilon)>20$~GeV and $|y(\Upsilon)|<1.2$ \cite{CMS:2026ccg} 
and constitutes a $\sim 3\sigma$ deviation from the Standard Model (SM) prediction of lepton flavor universality. Of note, results from $B$-factories \cite{PLUTO:1979xaz,DESY-Hamburg-Heidelberg-Munich:1980ttv,CLEO:1983iqm,ARGUS:1985eco,CLEO:1994orn,CLEO:2004tkr,CLEO:2006uhx} find the $\Upsilon$ decay width to $\tau^+\tau^-$ to be the same as that of the $\Upsilon \to \mu^+\mu^-$ channel (up to kinematic factors). 
 
The visible invariant mass of the scouted ditau events has a broad peak at $\sim 7$~GeV, consistent with the ditau spectrum expected given the masses of the three $\Upsilon$ states $1S$, $2S$, and $3S$ at 9.46, 10.0, and 10.3~GeV. Of these, the $\Upsilon(1S)$ production rate dominates in the Standard Model.

If the large ditau rate is interpreted not as the $\Upsilon$, but instead some other particle close in mass, the spin-0 $\eta_b$ (which has a mass of $9.399 \pm 0.002~\rm GeV$) seems at first to be a likely candidate.
However, the production times branching ratio into $\tau^+\tau^-$ final states is expected to be far too small to accommodate the signal from the SM decay of the $\eta_b$ \cite{Rashed:2010jp}. 

Taking all these facts together, the measured ditau rate seen at CMS seems difficult to accommodate in the SM.
 Even extended beyond the SM, it does not seem possible to reconcile the CMS result with the $B$-factory measurements by modifying the $\Upsilon(1S)$ leptonic branching ratios.

In this Letter, we consider a new physics scenario that can be accommodated by the data: that of an extended two-Higgs doublet model containing a pseudoscalar singlet (2HDM$+a$).
The relatively simple addition to the SM with an additional Higgs doublet and a singlet has been of interest to theorists as a component of the Next-to-Minimal Supersymmetric Standard Model (NMSSM, see Refs.~\cite{Maniatis:2009re,Ellwanger:2009dp} for reviews), as well as as a source of baryogenesis~\cite{Huber:2022ndk,Gent:2025csq} or portals to the dark sector \cite{Ipek:2014gua,No:2015xqa,Goncalves:2016iyg,Bauer:2017ota,Tunney:2017yfp,LHCDarkMatterWorkingGroup:2018ufk,Robens:2021lov,Arcadi:2022lpp}.

In this model, the ditau measurement arises from a light $\sim 9.5$~GeV pseudoscalar.
The state is predominantly the pseudoscalar singlet and couples to SM fermions through a small mixing with the heavy $SU(2)$ doublet pseudoscalar.
The mass-proportional Yukawa couplings suppress the branching ratio of the decay to muons while maintaining a sizable branching fraction into pairs of taus.
The $b$-quark decays of the singlet are both kinematically suppressed and swamped by the enormous hadronic backgrounds at low invariant mass. As a result, the ditau final states dominate the experimental signature. At the LHC, the pseudoscalar is produced primarily through a gluon-fusion process, via a loop of $b$-quarks. This model would have escaped detection at the $B$-factories due to the small coupling of the new pseudoscalar to electrons, and avoided constraints from LEP due to small couplings to the electroweak gauge bosons.

This model has an unlikely coincidence of scales between the light pseudoscalar and the $\Upsilon$. While no elegant solution to this serendipitous alignment of masses is currently available, it would provide an explanation for the prior absence of new physics signals at the LHC: the new particle would be effectively hiding in the shadow of the $\Upsilon$.

{\bf The 2HDM+$a$ model.}
We extend the Standard Model Higgs sector by introducing a second $SU(2)$ Higgs doublet \cite{Hall:1981bc,Gunion:2002zf,Branco:2011iw}, along with a real pseudoscalar $a_0$ which is a singlet under the SM gauge group.
As in the conventional 2HDM, we impose a discrete $\mathbb{Z}_2$ symmetry under which $\Phi_1$ is even, while $\Phi_2$ and $a_0$ are odd.
This symmetry eliminates flavor-changing neutral currents at tree level.
The 2HDM is the minimum viable Higgs sector in supersymmetry, though in this Letter we do not make any such additional assumptions. While several other configurations are possible, we adopt the Type-II 2HDM, in which $\Phi_2$ couples to the up-type quarks $u$, while $\Phi_1$ couples to the down-type quarks $d$ and the charged leptons $\ell$:
\begin{equation}
    {\cal L}_{\rm Yuk.} = y_u Q \Phi_2 \bar{u} + y_d Q \Phi_1^\dag  \bar{d}+ y_\ell L \Phi_1^\dag \bar{\ell}_R +{\rm h.c.}
\end{equation}
This structure can be enforce by assigning the right-handed up-type quarks to be odd under the $\mathbb{Z}_2$ symmetry.
Following the convention in Ref.~\cite{Gunion:2002zf,Gent:2025csq}, the scalar potential is given by:
\begin{eqnarray}
    V_{\rm 2HDM} & = & m_{11}^2 \Phi_1^\dagger\Phi_1 + m_{22}^2 \Phi_2^\dagger\Phi_2
     - m_{12}^2 \left[\Phi_1^\dagger \Phi_2 + \mathrm{h.c.}\right] \nonumber \\
    & & +\frac{1}{2}\lambda_1 |\Phi_1^\dagger \Phi_1|^2+ \frac{1}{2}\lambda_2 |\Phi_2^\dagger \Phi_2|^2 \nonumber \\
    & &  + \lambda_3 (\Phi_1^\dagger \Phi_1) (\Phi_2^\dagger \Phi_2) \\
    & & + \lambda_4 |\Phi_1^\dagger \Phi_2|^2 + \frac{\lambda_5}{2} \left[ (\Phi_1^\dagger \Phi_2)^2 + \mathrm{h.c.} \right]\,, \nonumber \\
    V_S & = & \frac{1}{2} \mu_a^2 a_0^2 + \frac{\lambda_a}{4} a_0^4 + \frac{\lambda_{1a}}{2} a_0^2 \Phi_1^\dagger \Phi_1 \nonumber \\
   & & + \frac{\lambda_{2a}}{2} a_0^2 \Phi_2^\dagger \Phi_2 + i \kappa a_0 (\Phi_1^\dagger \Phi_2 - \Phi_2^\dagger \Phi_1)\,,
\end{eqnarray}
where a soft $\mathbb{Z}_2$ breaking term $m_{12}^2$ is included as usual.
We work in the CP-conserving limit for simplicity, which requires $m_{12}^2$, $\lambda_5$, and $\kappa$ to be real.

After electroweak symmetry breaking, the two doublets can be expanded around their vacuum expectation values (vev) as
\begin{equation}
    \Phi_i =\left(\begin{array}{c} \phi^+ \\ (v_i+\rho_i +i \eta_i)/\sqrt{2} \end{array} \right).
\end{equation}
The Standard Model Higgs vev satisfies $v^2 = (246~{\rm GeV})^2 = v_1^2+v_2^2$, and the ratio of the vevs is defined as $\tan \beta \equiv v_2/v_1$.
For simplicity, we set $\langle a_0\rangle = 0$.
This is ensured by having positive $\mu_a^2$ and $\lambda_a$.
Requiring the vev to be the local minimum of the potential imposes the following conditions:
\begin{align}
    m_{11}^2 &= m_{12}^2 \tan \beta-\frac{1}{2} \lambda_{345} v_2^2-\frac{1}{2} \lambda _1 v_1^2,\label{eq:vev1} \\
    m_{22}^2 &= \frac{m_{12}^2}{\tan \beta} -\frac{1}{2} \lambda_{345} v_1^2-\frac{1}{2} \lambda _2 v_2^2\,,\label{eq:vev2}
\end{align}
where we have defined $\lambda_{345} \equiv \lambda_3 + \lambda_4 + \lambda_5$.
This potential also motivates our choice of a pseudoscalar over a scalar model: if $a_0$ is replaced by a scalar field, the linear term $\kappa$ becomes a tadpole term for the scalar, generating a non-zero vev. Given the resulting mass spectrum, we could not identify a benchmark point that maintains a light singlet scalar while keeping the doublets heavy enough to avoid the experimental bounds.

There are eight degrees of freedom in the 2HDM, and one additional degree from the pseudoscalar.
Three of them are Goldstone modes, while the remaining six result in one charged Higgs $H^\pm$, two neutral pseudoscalars, and two neutral scalars.
The physical charged Higgs $H^\pm$ acquires a mass %and neutral pseudoscalar $A$ acquire masses \MB{we didn't list the heavy $A$ mass here}
\begin{align}
    m_{H^\pm}^2 &= \left(\frac{m_{12}^2}{v_1 v_2} -\frac{\lambda_4 + \lambda_5}{2}\right) v^2\,,
\end{align}
The CP-even scalar sector is the same as the conventional 2HDM model, with a mass matrix
\begin{align}
    M^2_{\rm{even}} =
\begin{pmatrix}
\lambda_1 v_1^2 + m_{12}^2\tan\beta &
\lambda_{345} v_1 v_2 - m_{12}^2
\\
\lambda_{345} v_1 v_2 - m_{12}^2 &
\lambda_2 v_2^2 + \dfrac{m_{12}^2}{\tan\beta}
\end{pmatrix}\,.
\end{align}
The mass eigenstate is thus a mixture of $\rho_1$ and $\rho_2$, with
\begin{align}
\begin{pmatrix}
H\\
h
\end{pmatrix}
=
    \begin{pmatrix}
\cos\alpha & \sin\alpha
\\
-\sin\alpha & \cos\alpha
\end{pmatrix}
\begin{pmatrix}
\rho_1\\
\rho_2
\end{pmatrix}.
\end{align}
Meanwhile, the pseudoscalar sector is a mixture between $\eta_1$, $\eta_2$, and $a_0$.
After rotating the basis by $\beta$, one state becomes the neutral Goldstone boson to be eaten by the $Z$ boson, while the other two physical degrees of freedom have the  mass matrix
\begin{align}
M^2_{\text{odd}} =
\begin{pmatrix}
\left( \frac{m_{12}^2}{v_1 v_2} - \lambda_5 \right) v^2  & \kappa v \\
\kappa v & \mu_a^2 + \frac{1}{2} \left( \lambda_{a1} v_1^2 + \lambda_{a2} v_2^2 \right)
\end{pmatrix}\,.
\end{align}
Here, the corresponding gauge eigenstates are $A_0 = -\eta_1 \sin \beta + \eta_2 \cos \beta$ and $a_0$.

The interactions of these physical mass eigenstates with SM gauge bosons and fermions are determined by diagonalizing the mass matrix $M_{\rm odd}^2$:
\begin{equation}
\begin{pmatrix}
A_0\\
a_0
\end{pmatrix} =
\begin{pmatrix}
\cos\theta & \sin\theta \\
-\sin\theta & \cos\theta
\end{pmatrix}
\begin{pmatrix}
A\\
a
\end{pmatrix}.
\end{equation}

An experimentally viable model must live in a restricted corner of the possible parameter space to be consistent with experimental measurements of the Higgs sector. 
In particular, the SM-like $h$ must be largely decoupled from the singlet, and exist in the alignment limit $\beta - \alpha \approx \pi/2$ \cite{Gunion:2002zf,Haber:2013mia,Bernon:2015qea} of the two $\rho_i$ states.

\begin{table*}[t]
  \centering
  \begin{tabular*}{\textwidth}{@{\extracolsep{\fill}} c c c c c c c c c c c @{}}
    \hline\hline
    $\tan\beta$ &
    $m_{12}^2~[\mathrm{GeV}^2]$ &
    $\lambda_1$ &
    $\lambda_2$ &
    $\lambda_3$ &
    $\lambda_4$ &
    $\lambda_5$ &
    $\lambda_{1a}$ &
    $\lambda_{2a}$ &
    $\mu_a^2~[\mathrm{GeV}^2]$ &
    $\kappa~[\mathrm{GeV}]$
    \\[2pt]
    \hline
    50 & 91271.287 & 0.2583 & 0.2583 & 1.501 & 2.176 & -3.419 & -2.536 & -0.220 & 13201.011 & 709.904
    \\[2pt]
    \hline\hline
  \end{tabular*}
  \caption{Input parameters for the 2HDM+$a$ benchmark points (Type-II Yukawa structure) that are consistent with the ditau measurement. In addition, $m_{11}^2$ and $m_{22}^2$ are fixed by vev conditions in~\cref{eq:vev1,eq:vev2}. $\lambda_a$ is irrelevant in this discussion.}
  \label{tab:benchmark-input}
\end{table*}

\begin{table*}[t]
  \centering
  \begin{tabular*}{\textwidth}{@{\extracolsep{\fill}} c c c c c c@{}}
    \hline\hline
    $m_a$ [GeV] &
    $m_h$ [GeV] &
    $m_H$ [GeV] &
    $m_{H^\pm}$ [GeV]&
    $m_A$ [GeV]&
    $\theta$
    \\[2pt]
    \hline
    $9.5$ & $125$ & $2137$ & $2145$ & $2186$ & $-0.0366$
    \\[2pt]
    \hline\hline
  \end{tabular*}
  \caption{Masses and coupling ratios for the mass eigenstates which result from the parameter point shown in Table~\ref{tab:benchmark-input}. 
  }
  \label{tab:benchmark-eigenstates}
\end{table*}

The couplings of the neutral states $X = H, A, a$ to quarks and leptons $i=u,d,\ell$ are proportional to the corresponding fermion masses, 
and are typically expressed in terms of the ratios $\kappa_X^i$ of the couplings to the Standard Model Higgs Yukawas.
For the Type-II 2HDM adopted in this work, our convention gives
\begin{align}
    \label{eq:kappas}
    &\kappa^\ell_{a} = \kappa^d_{a} = \tan \beta \sin \theta, ~~ \kappa_a^u = \cot \beta \sin \theta\,,\nonumber \\
    &\kappa^\ell_{A} = \kappa^d_{A} = \tan \beta \cos \theta, ~~ \kappa_A^u = \cot \beta \cos \theta\,,\nonumber \\
    &\kappa_H^\ell = \kappa_H^d = \tan \beta, ~~\kappa_H^u = - \cot \beta\,.
\end{align}
In the alignment limit, the SM Higgs coupling is not modified, and the exotic neutral states do not couple to SM gauge bosons. For simplicity, we assume exact alignment so that the $a$ coupling to SM gauge bosons vanishes. As a result, the light pseudoscalar evades both electroweak precision test~\cite{Arcadi:2022lpp,Li:2025zga} bounds and the $Za$ production bound at LEP~\cite{L3:1996ome}.

The charged $H^\pm$ are difficult to significantly separate in mass from the heavy $H$. These charged states must be above $\sim 1400$~GeV for large $\tan \beta$ \cite{ATLAS:2024hya}; this requires us to identify the $h$ (rather than the $H$) as the 125~GeV Standard Model-like Higgs discovered by CMS and ATLAS. The most general scalar sector has a wide range of possible phenomenology; however we are constrained by the dual requirements that the 125~GeV $h$ must be Standard Model-like and the $\sim 9.5$~GeV $a$ must be predominantly singlet with enough mixing to the doublets to be produced in sufficient quantities and decay to $\tau^+\tau^-$.

{\bf Parameter choice and experimental signals.}
In this Letter, we select a single example parameter point at the alignment limit that contains a pseudoscalar $a$ with the appropriate cross section and branching ratios to explain the CMS ditau measurement, while satisfying all existing experimental constraints on an extended Higgs sector.
Our goal here is to provide an existence proof that such an explanation is viable; a full exploration of the parameter space is left to future work~\cite{Buckley:2026prep}.

Table~\ref{tab:benchmark-input} shows the Lagrangian parameters for our working point, while 
the resulting mass spectrum and coupling ratios are provided in Table~\ref{tab:benchmark-eigenstates}.
Our parameter point balances the small mixing of the light singlet pseudoscalar $a$ with the doublet pseudoscalar $A$ with a large $\tan\beta$, resulting in a significant coupling to the down-type quarks and leptons. 

The light pseudoscalar $a$ is -- by design -- close to the mass of the $\Upsilon(1S)$.
As a result, its decays into tau final states are kinematically indistinguishable from those of the Standard Model resonance, and would be interpreted by the CMS analysis as an increase in the cross section times branching ratio of $\Upsilon(1S) \to \tau^+\tau^-$.

Since $m_a < 2 m_B$ where $m_B$ is the lightest B meson mass, the decay into $\bar{b}b$ is kinematically forbidden.
Its decays are thus likewise primarily to $\tau$ leptons and gluon pairs through the $b$-loop.
The branching ratios to these final states are:
\begin{align}
    &{\rm BR}(a\to \tau^+\tau^-) =0.875 ,\nonumber \\
    &{\rm BR}(a\to gg) = 0.12\,,
\end{align}
where we used the formula calculated in Refs.~\cite{Djouadi:2005gj,Hooper:2025iii}.
 The branching ratio to muons is suppressed relative to that of taus by the ratio of the mass-squared:
\begin{equation}
    \frac{\Gamma(a\to \mu^+\mu^-)}{\Gamma(a\to \tau^+\tau^-)} =\frac{m_\mu^2}{m_\tau^2} \approx \frac{1}{280}.
\end{equation}
    
We estimate the production cross section using \texttt{MadGraph v3.6.6} \cite{Alwall:2014hca,Frederix:2018nkq}, using our customized version of the model files for general 2HDM~\cite{Branco:1999fs,Degrande:2014qga} to include a pseudoscalar and the built-in 5-flavor-structure parton distribution function. We then run \texttt{Pythia v8.316}~\cite{Bierlich:2022pfr} to simulate the initial-state radiation, which transversely boosts the produced $a$. We apply cuts of $|\eta| < 1.2$ and $p_T > 20$ GeV on $a$ to acquire the fiducial cross section.
At $\sqrt{s}=13.6$~TeV, we find the cross section in the fiducial region is dominated by $\bar{b}b \to a$ with a cross section of 3.26 nb.
The contribution in the fiducial region from gluon fusion via the $b$-loop is 0.37 nb.
The total production cross section in the fiducial region is 3.65 nb, which becomes 3.2 nb after the branching fraction into taus.

Such an explanation for the ditau events can be consistent with the measurement of the dimuon channel~\cite{CMS:2026ccg}.
We assume the acceptance for muons from the $a$ decay
in the analysis window is similar to that of the $\Upsilon(1S)$ decay.
From the ditau measurement, the production cross section times branching ratio into taus of $a$ is $\sim 17\pm7$ times that of the $\Upsilon(1S)$.
However, while the branching ratio of the $\Upsilon(1S)$ into muons is the same as that into taus, the $a$ branching ratio to muons is only $0.35\%$ of taus.
As a result, the $a$ contribution to the dimuon resonance near $9.5$~GeV  is $\sim (6\pm3)\%$ of that $\Upsilon(1S) \to \mu^+\mu^-$.
Ref.~\cite{CMS:2026ccg} identifies systematic uncertainties in the measured cross section in the muon channel that vary from $3-10\%$ across the $p_T$ range of the analysis.
It appears then that the $a$ contribution to $\Upsilon(1S)$ peak in dimuons could then be absorbed by the uncertainties in that analysis.

As far as other experimental constraints are concerned, our benchmark is consistent with the current limits on  the heavy Higgs: $\gtrsim 1950$~GeV from neutral Higgs direct production~\cite{ATLAS:2020zms,CMS:2022goy}, $\gtrsim 1400$~GeV from $p p \to t b H$ production~\cite{ATLAS:2024hya} and $\gtrsim 800$~GeV from $B \to X \gamma$ decay~\cite{HFLAV:2016hnz,Misiak:2017bgg}. 
This model predicts the Higgs exotic decay $h \to aa$ with a branching ratio of $4.97 \times 10^{-4}$, well below the current sensitivity of $3 \times 10^{-2}$~\cite{ATLAS:2025qyn}.

{\bf Conclusion and discussions.} We present a simple extension of the Standard Model that provides a possible explanation for the $\sim 3\sigma$ upward fluctuation in the $\Upsilon \to \tau^+\tau^-$ rate measured by CMS, introducing an extended Higgs sector and a singlet pseudoscalar which is coincidentally close to the $\Upsilon(1S)$ mass. Through mixing with Higgs doublets, the primarily-singlet new particle at $\sim 9.5$~GeV inherits Yukawa-proportional couplings, explaining the enhanced ditau rate as compared to the dimuon final states. Placing the 2HDM in the alignment limit eliminates couplings to electroweak gauge bosons; as a result the new particle would not have been produced at high rates in previous electron colliders.

The measurement at CMS is an intriguing result in an extremely difficult channel, and will face an immediate test in the $\sim 250$~fb$^{-1}$ of CMS data taken between 2024 and 2026. ATLAS has a similar amount of data, and doubtlessly can perform an independent test in low-$p_T$ ditaus.
If the statistical significance of the ditau rate increases, the 2HDM+$a$ model predicts rich phenomenology which would distinguish this scenario from other possibilities.
The most promising way to probe our benchmark point is via the Higgs exotic decay $h \to aa$, 
which is within the future sensitivity of the High Luminosity LHC~\cite{Carena:2022yvx} and future $e^+ e^-$ collider~\cite{Liu:2016zki}.
Heavy scalars $H/A/H^{\pm}$ can be produced at colliders associated with $b$ or $t$ quarks, or lead to anomalous $B \to X \gamma$ decay.

{\bf Acknowledgements.} 
We thank Thomas Biekotter for useful discussions.
I.R.W.\, is supported by Fermi Forward Discovery Group, LLC under Contract No. 89243024CSC000002 with the U.S. Department of Energy, Office of Science, Office of High Energy Physics.
I.R.W.\, is also supported by DOE distinguished scientist fellowship grant FNAL 22-33.
D.S.\, and M.R.B.\, are supported by the DOE under Award Number DOE-SC0010008.

\bibliography{main}

\end{document}